\documentclass[aps,pra,12pt,showkeys,superscriptaddress,showpacs,reprint,raggedbottom]{revtex4-1}
\usepackage{amsmath}
\usepackage{amsthm}
\usepackage{graphicx}
\usepackage[colorlinks]{hyperref}
\allowdisplaybreaks
\begin{document}

\title
{
Variational solution of congruent transformed Hamiltonian for many-electron systems using full configuration interaction calculation
}

\author{Jennifer M. Elward}
\affiliation{%
Department of Chemistry, Syracuse University, \\
Syracuse, New York 13244, USA
}
\author{Johannes Hoja}
\affiliation{%
Department of Chemistry,
University of Graz, \\
8010 Graz, Austria
}
\author{Arindam Chakraborty}
\email[corresponding author: ]{archakra@syr.edu}
\affiliation{%
Department of Chemistry, Syracuse University, \\
Syracuse, New York 13244, USA
}

\date{\today}

\begin{abstract}
The congruent transformation of the electronic Hamiltonian
is developed to address the electron correlation problem in many-electron
systems. 
The central strategy presented in this method is to
perform transformation on the electronic Hamiltonian 
for approximate removal of the Coulomb singularity. 
The principle difference between 
the present method and the transcorrelated method of Handy and Boys 
is that the congruent transformation preserves the Hermitian property of the Hamiltonian. 
The congruent transformation is carried out using explicitly correlated functions
and the optimum 
correlated transform Hamiltonian is obtained by performing a search over a set of
transformation functions.  
The ansatz of the transformation functions are selected to facilitate
analytical evaluation of all the resulting integrals.
 The ground state energy is obtained variationally 
by performing a full configuration interaction (FCI) calculation 
on the congruent transformed Hamiltonian.
Computed results on well-studied benchmark systems
show that for the identical basis functions, the 
energy from the congruent transformed Hamiltonian is significantly lower
than the conventional FCI calculations. Since the number of configuration
state functions in the FCI calculation increases rapidly with the size of the 1-particle basis
set, the results indicate that the congruent transformed
Hamiltonian provides a viable alternative to obtain FCI quality energy using a 
smaller underlying 1-particle basis set. 
\end{abstract}


 %Use showkeys class option if keyword
 %display desired

\maketitle

\section{Introduction}
The form of the many-electron wavefunction in the proximity of the  
electron-electron and electron-nuclear coalescence  
point plays a  critical role in accurate determination  
of the ground and excited state energies.
Although, the precise structure of the many-electron wavefunction 
continues to be  elusive, the form of the exact wavefunction at the coalescence point is well understood and is given by the Kato cusp condition.~\cite{CPA:CPA3160100201,hammond1994monte,lester1997recent,nightingale1999quantum}
In the many-electron wavefunction, 
the electron-nuclear cusp condition can be incorporated by using Slater-type orbitals (STOs). 
For calculations involving Gaussian-type orbitals (GTOs), 
the one-electron basis can be improved iteratively by adding GTOs
with increasing angular momentum quantum number.\cite{Klopper1986339} The subject of convergence
of single-particle basis has
been analyzed extensively using both analytical and numerical techniques.~\cite{CuevasSaavedra2012163,papajak:064110,yanai:084107}
The electron-electron cusp has been the focus of intense research 
because of its direct relation to the electron correlation problem
and accurate description of the Coulomb and Fermi hole.~\cite{Frolov2011305,PhysRevA.72.062502,QUA:QUA22109,prendergast:1626,shiozaki:141103}
However, unlike the 
electron-nuclear cusp, atom-centered basis functions are not ideal for accurate 
description of the many-electron wavefunction near the electron-electron 
cusp.~\cite{doi:10.1021/cr200204r,hattig_review,clark:244105} Indeed it has been shown that the slow convergence of a 
full configuration interaction (FCI) calculation with respect to the 
one-particle basis is related to the inadequate treatment of the 
electron-electron cusp.~\cite{hattig_review} The solution 
is to include explicit $r_{12}$ dependence in the form of the wavefunction,
and there is a large assortment of quantum chemical methods that  
have incorporated this approach. For example, in the variational Monte Carlo (VMC) method, the Jastrow function is used for
including explicit $r_{12}$ terms in the trial wavefunction.~\cite{nightingale1999quantum,hammond1994monte} 
The form of the Jastrow is chosen to ensure that the 
electron-electron and electron-nuclear Kato cusp conditions are satisfied. 
The parameters in the Jastrow function are 
obtained by minimizing the linear 
combination of energy and its variance.
Because of the complicated mathematical form of the Jastrow function it is not
possible to evaluate the integral over the electronic coordinates analytically, and 
a stochastic numerical method is used for computation of the energy. 
Recently, Morales \emph{et al.}  performed highly
accurate multi-determinant VMC calculations on water.~\cite{doi:10.1021/ct3003404}
 A detailed 
review of various applications of Quantum Monte Carlo (QMC) methods in physics and 
chemistry can be found in Ref. \onlinecite{doi:10.1021/cr2001564,Ceperley01012010,RevModPhys.73.33}.
Explicitly correlated methods have also been developed for 
post Hartree-Fock schemes such as 
perturbation theory (MP2-R12), coupled-cluster methods (CC-R12),
and multireference CI schemes (R12-MRCI).
These methods introduce the electron-hole interparticle distance 
directly into the calculation in order to increase the accuracy of the calculations. 
The field of explicitly correlated methods for electronic 
structure calculation has been reviewed and a
detailed description of various methods can be found in 
Ref. \onlinecite{doi:10.1021/cr200204r,hattig_review,springerlink:10.1007/s00214-011-1070-1}.
A common feature of the R12 and F12 methods discussed above is that
they all involve analytical computation of the  $r_{12}$ correlation function. 
Recently, Chinnamsetty and coworkers have presented an interesting study that 
compared and contrasted QMC with various F12 methods.~\cite{Chinnamsetty201236}

A different strategy known as the transcorrelated method was developed
by Handy and Boys in 1969.~\cite{Boys29041969}
The basic idea of the transcorrelated
method is to remove the electron-electron Coulomb singularity 
by performing similarity transformation 
on the Hamiltonian using an explicitly correlated function.
The method was later extended by Ten-no to treat the electron-electron 
cusp using Guassian geminal functions and was applied to chemical systems.~\cite{Tenno2000169,yanai:084107}
The transcorrelated method has also been combined with other methods such as QMC~\cite{QUA:QUA10455} and coupled-cluster theory~\cite{Hino2002317} 
and has been used to study electron correlation in periodic systems.~\cite{PhysRevB.69.165102}  
One of the defining characteristics of this method is that the transcorrelated Hamiltonian
is not Hermitian and therefore is not required to be bounded from below by the exact ground state energy. 
The correlation function can be obtained either by minimizing the energy variance of the 
transcorrelated Hamiltonian~\cite{doi:10.1080/00268977100101961,doi:10.1080/00268977200100011} or by requiring the correlation function to satisfy
the electron-electron cusp condition.

The focus of the present work is to address the non-Hermitian property of the transcorrelated 
Hamiltonian by replacing the similarity transformation by congruent transformation.~\cite{james1968mathematics,datta2004matrix,watkins2002fundamentals}
By performing 
congruent transformation, we preserve the Hermitian property of the electronic Hamiltonian
which allows us to use a standard electronic structure method such as configuration interaction method
to minimize the total energy. The remainder of the paper describes the theoretical development and the implementation details of the method. The derivation of the 
congruent transformed Hamiltonian is presented in Sec. \ref{sec:cth}. Details of 
performing FCI calculations using the congruent transformed Hamiltonian and interfacing it with existing FCI methods are presented in Sec. \ref{sec:corrfunc} and
\ref{sec:trialwf}.
Benchmark calculations using the congruent transformed Hamiltonian are
presented in Sec.~\ref{sec:comp_details}. The analysis of the results and conclusions are presented in Sec. \ref{sec:discussion_conclusions}.

\section{Congruent transformed Hamiltonian}
\label{sec:cth} 
The congruent transformed (CT) Hamiltonian $\tilde{H}$ is defined by performing the 
following transformation \cite{james1968mathematics,datta2004matrix,watkins2002fundamentals}
\begin{align}
	\tilde{H} = G^\dagger H G, 
\end{align}
where $G$ is an explicitly correlated function which will be defined later. 
The expectation value of the CT Hamiltonian with respect to any trial wavefunction is given as
\begin{align}
	\tilde{E}_\mathrm{T}[\Psi_\mathrm{T},G]
	=
	\frac
	{\langle \Psi_\mathrm{T} \vert \tilde{H} \vert \Psi_\mathrm{T} \rangle} 
	{\langle \Psi_\mathrm{T} \vert \tilde{1} \vert \Psi_\mathrm{T} \rangle},
\end{align}
where $\tilde{1} = G^{\dagger}1G$. The above expression is mathematically equivalent to calculating the expectation  value of 
the electronic Hamiltonian using a correlated wavefunction and is bounded from below by the
exact ground state energy $E_\mathrm{exact} \leq \tilde{E}_\mathrm{T}$.
The optimized energy associated with the CT Hamiltonian is obtained by performing a minimization with respect to the trial wavefunction and explicitly correlated function,
\begin{align}
	E_\mathrm{CT} 
	&=
    \min_{\Psi_\mathrm{T}} \min_{G}   \tilde{E}_\mathrm{T}[\Psi_\mathrm{T},G].
\end{align} 

The optimization of the correlation function $G$ and the trial wavefunction $\Psi_\mathrm{T}$
is conducted in two steps. In the first step, the form of the trial function is kept
fixed to a single Slater determinant and the parameters of the geminal functions
are determined by minimizing the geminal parameters and the molecular  orbitals. In the
second step, the minimized geminal function $G_\mathrm{min}$ is kept fixed and the 
trial wavefunction $\Psi_\mathrm{T}$ is minimized. The steps involved are described by the following 
equation
\begin{align}
\label{eq:gmin}
	\tilde{E}[G_\mathrm{min}]
	&=
	\min_{G,\Phi_\mathrm{SD}} 
	\tilde{E}_\mathrm{T}[\Phi_\mathrm{SD},G], \\
\label{eq:ect}
	E_\mathrm{CT} 
	&=
	\min_{\Psi_\mathrm{T}}  \tilde{E}_\mathrm{T}[\Psi_\mathrm{T},G_\mathrm{min}].
\end{align}
The optimization of the correlation function and the trial wavefunction 
are described in the following subsections.

\subsection{Optimization of the correlation function}
\label{sec:corrfunc}
The choice of the correlation function $G$ plays an important part in the 
implementation of the method for practical applications. In principle, 
a variety of correlated functions such as two and three-body Jastrow functions 
can be used. However, the matrix elements associated with these functions
cannot be integrated analytically and one has to use numerical techniques
such as the VMC method to calculate the integrals. 
In the present work, Gaussian-type geminal (GTG) functions were used for the correlated functions. 
The GTG functions were introduced by Boys \cite{Boys22021950,Boys25101960} and Singer \cite{Singer25101960}, and have
been used extensively in explicitly correlated methods.~\cite{komornicki:244115,korona:5109,bukowski:3306,bukowski:4165,chakraborty:014101,Refworks:204} 
Slater determinants augmented with GTG functions have been used to 
study electron-electron and electron-proton systems.
The integrals involving
GTG functions with GTOs can be performed analytically and have been derived earlier.~\cite{persson:5915,springerlink:10.1007/s002140050258,Boys22021950,Boys25101960}
The form of the correlated function used in the following calculations is defined 
as
\begin{align}
	G = \sum_{i<j}^{N} g(i,j),
\end{align}
\begin{align}
	g(i,j) = \sum_{k=1}^{N_\mathrm{g}} b_k e^{-\gamma_k r_{ij}^2},
\end{align}
where $N$ is number of electrons and $N_\mathrm{g}$ is the number of Gaussian functions.
The geminal coefficients $\{b_k,\gamma_k \}$ in the GTG function are determined 
variationally. 
 In the limit of $G \rightarrow 1 $, the energy $\tilde{E}[G_\mathrm{min}] $  becomes equal to the Hartree-Fock energy.
\begin{align}
	E_{\mathrm{HF}} = \lim_{G \rightarrow 1} \tilde{E}[G_\mathrm{min}],
\end{align} 
As a consequence, the HF energy is the upper bound to the geminal minimization 
process
\begin{align}
	\tilde{E}[G_\mathrm{min}] \leq E_{\mathrm{HF}}.
\end{align}
The transformed Hamiltonian is expanded as a sum of 2-6 particle operators as shown below
\begin{align}
\label{eq:O_operatots}
	\tilde{H}
	&=  \sum_{i<j} \sum_{k} \sum_{m<n} g(m,n) h_1(k) g(i,j) \\ \notag
	&+ \sum_{i<j} \sum_{k<l} \sum_{m<n} g(i,j) r_{kl}^{-1} g(m,n), \\
	&= O_2 + O_3 + O_4 + O_5 + O_6
\end{align}
where, the operators $\{O_n, n=2,\dots,6 \}$ are defined by collecting all two, three, four,
five and six particle operators obtained by expanding the summation in Eq. \eqref{eq:O_operatots}. Specifically,
\begin{align}
	O_2 &=  \sum_{i < j }  h_2(i,j), \\
	O_3 &=  \sum_{i < j < k }  h_3(i,j,k), \\
	O_4 &=  \sum_{i < j < k < l}  h_4(i,j,k,l), \\
	O_5 &=  \sum_{i < j < k < l < m}  h_5(i,j,k,l,m), \\ 
	O_6 &=  \sum_{i < j < k < l < m < n}  h_6(i,j,k,l,m,n).
\end{align} 
The exact form of the operators $\{h_n, n=2,\dots,6 \}$ have been derived earlier and are not duplicated here.~\cite{elward:124105}
It should be emphasized that
the operators $\{h_n, n=2,\dots,6 \}$
are defined so that they are completely symmetric with respect to 
all $n!$ permutation of the indices
\begin{align}
	\mathcal{P}_k h_n = h_n \quad \mathrm{where} \, \mathcal{P}_k \in S_n.
\end{align}
The operator $\mathcal{P}_k$ is the permutation operator that belongs to 
the complete symmetric group $S_n$. An important feature of this method is 
the availability of the analytical gradients of the total energy with 
respect to the geminal parameters. The gradients can be computed analytically and are given by the following expressions
\begin{align}
	\frac{\partial g(1,2) }{\partial b_k} &= e^{-\gamma_k r_{12}^2}, \\
	\frac{\partial g(1,2) }{\partial \gamma_k} &= -b_k r_{12}^2 e^{-\gamma_k r_{12}^2}.
\end{align}
The AO integrals involving the gradients of the GTG 
functions are performed analytically and are computed with other AO integrals.

\subsection{Optimization of the trial wavefunction}
\label{sec:trialwf}
The optimization of the trial wavefunction $\Psi_\mathrm{T} $
is performed by performing a full configuration interaction (FCI) calculation on the 
congruent transformed Hamiltonian. The FCI wavefunction
is constructed by performing all possible excitations from the 
reference wavefunction.~\cite{szabo} This can be represented by the following expression,
\begin{align}
	\Psi_{\mathrm{FCI}} 
	&=
	C_0 \Phi + \sum_{a}^{N_\mathrm{occ}} \sum_{p}^{N_\mathrm{vir}} C_{a}^{p} {\Phi}_{a}^{p}
	+ \sum_{a<b}^{N_\mathrm{occ}} \sum_{p<q}^{N_\mathrm{vir}} C_{ab}^{pq} {\Phi}_{ab}^{pq} \\ \notag
	&+\sum_{a<b<c}^{N_\mathrm{occ}} \sum_{p<q<r}^{N_\mathrm{vir}} C_{abc}^{pqr} {\Phi}_{abc}^{pqr}
	+ \dots, 
\end{align}
where we have retained $N_\mathrm{vir}$ in the 
expression to emphasize that only a finite number of 
terms are evaluated.  This point will be a subject 
of discussion later in the derivation. The occupied and virtual orbitals are represented by $(a,b,c,\dots)$ and $(p,q,r,\dots)$, respectively,
and the CI coefficients are represented by 
$(C_a^p,\dots)$ and are obtained variationally by minimizing the total energy. The construction of the full set of excitations
and the determination of the CI coefficients are the two principle computational challenges associated with the FCI method. For very small molecules, the CI matrix can be explicitly
constructed and diagonalized, however, this simple
approach becomes prohibitively expensive as the system size increases.
Currently, there are various computational techniques for efficient calculation of 
the expansion coefficients.~\cite{Bak2001375,PhysRevB.81.115323,boguslawski:224101,PhysRevB.67.085314,prendergast:1626,rontani:124102}
The calculation requires matrix elements involving the operators $\{\langle \Phi_k \vert O_\alpha \vert \Phi_{k'} \rangle,\alpha=2,\dots,6\}$ 
which are derived below.

The matrix elements involving the 2-particle operators are evaluated as
\begin{align}
\label{eq:o2}
	\langle \Phi_0 \vert O_2 \vert \Phi_0 \rangle 
	&=
	\frac{1}{2!}
	\sum_{k=1}^{2!}
	\sum_{i_1 i_2}^{N_\mathrm{occ}}
	(-1)^{p_k}
    \langle i_1 i_2 \vert h_2 \vert 
    P_k i_1 i_2 \rangle , \\ 
	\langle \Phi_0 \vert O_2 \vert \Phi_{a}^{p} \rangle 
	&=
	\sum_{k=1}^{2!}
	\sum_{i_1}^{N_\mathrm{occ}}
	(-1)^{p_k}
    \langle a i_1 \vert h_2 \vert 
    P_k p i_1 \rangle , \\
	\langle \Phi_0 \vert O_2 \vert \Phi_{ab}^{pq} \rangle 
	&=
	\sum_{k=1}^{2!}
	(-1)^{p_k}
    \langle a b \vert h_2 \vert 
    P_k p q \rangle  .
\end{align}

The matrix elements involving the 3-particle operators are evaluated as
\begin{align}
\label{eq:o3}
	\langle \Phi_0 \vert O_3 \vert \Phi_0 \rangle 
	&=
	\frac{1}{3!}
	\sum_{k=1}^{3!}
	\sum_{i_1 i_2 i_3}^{N_\mathrm{occ}}
	(-1)^{p_k}
    \langle i_1 i_2 i_3 \vert h_3 \vert 
    P_k i_1 i_2 i_3 \rangle , \\ 
	\langle \Phi_0 \vert O_3 \vert \Phi_{a}^{p} \rangle 
	&=
	\frac{1}{2!}
	\sum_{k=1}^{3!}
	\sum_{i_1 i_2}^{N_\mathrm{occ}}
	(-1)^{p_k}
    \langle a i_1 i_2 \vert h_3 \vert 
    P_k p i_1 i_2 \rangle , \\
	\langle \Phi_0 \vert O_3 \vert \Phi_{ab}^{pq} \rangle 
	&=
	\sum_{k=1}^{3!}
	\sum_{i_1}^{N_\mathrm{occ}}
	(-1)^{p_k}
    \langle a b i_1 \vert h_3 \vert 
    P_k p q i_1 \rangle , \\
	\langle \Phi_0 \vert O_3 \vert \Phi_{abc}^{pqr} \rangle 
	&=
	\sum_{k=1}^{3!}
	(-1)^{p_k}
    \langle a b c \vert h_3 \vert 
    P_k p q r \rangle  .
\end{align}

The matrix elements involving the 4-particle operators are evaluated as
\begin{align}
\label{eq:o4}
	\langle \Phi_0 \vert O_4 \vert \Phi_0 \rangle 
	&=
	\frac{1}{4!}
	\sum_{k=1}^{4!}
	\sum_{i_1 i_2 i_3 i_4}^{N_\mathrm{occ}} \\ \notag
	& (-1)^{p_k}
    \langle i_1 i_2 i_3 i_4 \vert h_4 \vert 
    P_k i_1 i_2 i_3 i_4 \rangle , \\ 
	\langle \Phi_0 \vert O_4 \vert \Phi_{a}^{p} \rangle 
	&=
	\frac{1}{3!}
	\sum_{k=1}^{4!}
	\sum_{i_1 i_2 i_3}^{N_\mathrm{occ}}
	(-1)^{p_k}
    \langle a i_1 i_2 i_3 \vert h_4 \vert 
    P_k p i_1 i_2 i_3 \rangle , \\
	\langle \Phi_0 \vert O_4 \vert \Phi_{ab}^{pq} \rangle 
	&=
	\frac{1}{2!}
	\sum_{k=1}^{4!}
	\sum_{i_1 i_2}^{N_\mathrm{occ}}
	(-1)^{p_k}
    \langle a b i_1 i_2 \vert h_4 \vert 
    P_k p q i_1 i_2 \rangle , \\
	\langle \Phi_0 \vert O_4 \vert \Phi_{abc}^{pqr} \rangle 
	&=
	\sum_{k=1}^{4!}
	\sum_{i_1}^{N_\mathrm{occ}}
	(-1)^{p_k}
    \langle a b c i_1 \vert h_4 \vert 
    P_k p q r i_1 \rangle , \\
    \langle \Phi_0 \vert O_4 \vert \Phi_{abcd}^{pqrs} \rangle 
	&=
	\sum_{k=1}^{4!}
	(-1)^{p_k}
    \langle a b c d \vert h_4 \vert 
    P_k p q r s \rangle .
\end{align}

The matrix elements involving the 5-particle operators are evaluated as
\begin{align}
\label{eq:o5}
	\langle \Phi_0 \vert O_5 \vert \Phi_0 \rangle 
	&=
	\frac{1}{5!}
	\sum_{k=1}^{5!}
	\sum_{i_1 i_2 i_3 i_4 i_5}^{N_\mathrm{occ}} \\ \notag
	& (-1)^{p_k}
    \langle i_1 i_2 i_3 i_4 i_5 \vert h_5 \vert 
    P_k i_1 i_2 i_3 i_4 i_5 \rangle , \\ 
	\langle \Phi_0 \vert O_5 \vert \Phi_{a}^{p} \rangle 
	&=
	\frac{1}{4!}
	\sum_{k=1}^{5!}
	\sum_{i_1 i_2 i_3 i_4}^{N_\mathrm{occ}} \\ \notag
	& (-1)^{p_k}
    \langle a i_1 i_2 i_3 i_4 \vert h_5 \vert 
    P_k p i_1 i_2 i_3 i_4 \rangle , \\
	\langle \Phi_0 \vert O_5 \vert \Phi_{ab}^{pq} \rangle 
	&=
	\frac{1}{3!}
	\sum_{k=1}^{5!}
	\sum_{i_1 i_2 i_3}^{N_\mathrm{occ}}
	(-1)^{p_k}
    \langle a b i_1 i_2 i_3 \vert h_5 \vert 
    P_k p q i_1 i_2 i_3 \rangle , \\
	\langle \Phi_0 \vert O_5 \vert \Phi_{abc}^{pqr} \rangle 
	&=
	\frac{1}{2!}
	\sum_{k=1}^{5!}
	\sum_{i_1 i_2}^{N_\mathrm{occ}}
	(-1)^{p_k}
    \langle a b c i_1 i_2 \vert h_5 \vert 
    P_k p q r i_1 i_2 \rangle , \\
    \langle \Phi_0 \vert O_5 \vert \Phi_{abcd}^{pqrs} \rangle 
	&=
	\sum_{k=1}^{5!}
	\sum_{i_1}^{N_\mathrm{occ}}
	(-1)^{p_k}
    \langle a b c d i_1 \vert h_5 \vert 
    P_k p q r s i_1 \rangle , \\
    \langle \Phi_0 \vert O_5 \vert \Phi_{abcde}^{pqrst} \rangle 
	&=
	\sum_{k=1}^{5!}
    \langle a b c d e \vert h_5 \vert 
    P_k p q r s t \rangle .      
\end{align}

The matrix elements involving the 6-particle operators are evaluated as
\begin{align}
	\langle \Phi_0 \vert O_6 \vert \Phi_0 \rangle 
	&=
	\frac{1}{6!}
	\sum_{k=1}^{6!}
	\sum_{i_1 i_2 i_3 i_4 i_5 i_6}^{N_\mathrm{occ}} \\ \notag
	& (-1)^{p_k}
    \langle i_1 i_2 i_3 i_4 i_5 i_6 \vert h_6 \vert 
    P_k i_1 i_2 i_3 i_4 i_5 i_6 \rangle , \\ 
	\langle \Phi_0 \vert O_6 \vert \Phi_{a}^{p} \rangle 
	&=
	\frac{1}{5!}
	\sum_{k=1}^{6!}
	\sum_{i_1 i_2 i_3 i_4 i_5}^{N_\mathrm{occ}} \\ \notag
	& (-1)^{p_k}
    \langle a i_1 i_2 i_3 i_4 i_5 \vert h_6 \vert 
    P_k p i_1 i_2 i_3 i_4 i_5 \rangle , \\
	\langle \Phi_0 \vert O_6 \vert \Phi_{ab}^{pq} \rangle 
	&=
	\frac{1}{4!}
	\sum_{k=1}^{6!}
	\sum_{i_1 i_2 i_3 i_4}^{N_\mathrm{occ}} \\ \notag
	& (-1)^{p_k}
    \langle a b i_1 i_2 i_3 i_4 \vert h_6 \vert 
    P_k p q i_1 i_2 i_3 i_4 \rangle , \\
	\langle \Phi_0 \vert O_6 \vert \Phi_{abc}^{pqr} \rangle 
	&=
	\frac{1}{3!}
	\sum_{k=1}^{6!}
	\sum_{i_1 i_2 i_3}^{N_\mathrm{occ}} \\ \notag
	& (-1)^{p_k}
    \langle a b c i_1 i_2 i_3 \vert h_6 \vert 
    P_k p q r i_1 i_2 i_3 \rangle , \\
    \langle \Phi_0 \vert O_6 \vert \Phi_{abcd}^{pqrs} \rangle 
	&=
	\frac{1}{2!}
	\sum_{k=1}^{6!}
	\sum_{i_1 i_2}^{N_\mathrm{occ}} \\ \notag
	& (-1)^{p_k}
    \langle a b c d i_1 i_2 \vert h_6 \vert 
    P_k p q r s i_1 i_2 \rangle , \\
    \langle \Phi_0 \vert O_6 \vert \Phi_{abcde}^{pqrst} \rangle 
	&=
	\sum_{k=1}^{6!}
	\sum_{i_1}^{N_\mathrm{occ}}
    \langle a b c d e i_1 \vert h_6 \vert 
    P_k p q r s t i_1 \rangle , \\
    \langle \Phi_0 \vert O_6 \vert \Phi_{abcdef}^{pqrstu} \rangle 
	&=
	\sum_{k=1}^{6!}
    \langle a b c d e f \vert h_6 \vert 
    P_k p q r s t u \rangle  .      
\end{align}
The computation of matrix elements in the above expression requires atomic orbital
integrals involving the GTG functions. One of the advantages of using the GTG functions
is that all the AO integrals needed for the CT Hamiltonian calculation can be computed 
analytically. Boys and Singer  have derived the integrals involving GTG functions with 
s-type GTOs. Persson and Taylor have extended the method for higher angular momentum by using the Hermite Gaussian expansion approach.~\cite{springerlink:10.1007/s002140050258}
 Recently, Hofener and coworkers 
have also derived the geminal integrals by extending the Obara-Saika techniques for 
calculating the GTG integrals.~\cite{Hofener200925} 

The solution for the CI coefficients requires diagonalization of the CI Hamiltonian 
matrix. However, the lowest eigenvalue and eigenfunction can
be obtained without explicit construction and storage of the CI matrix. There are 
various efficient methods such as the Davidson diagonalization to perform this task.~\cite{Davidson197587} Recently, Alavi et al. have developed the 
FCIQMC method which allows very efficient evaluation of the 
FCI wavefunction.~\cite{booth:054106,booth:174104,booth:084104,cleland:024112,shepherd:244101}

In the present calculation, the FCI eigenvector was obtained by performing the 
Nesbet update scheme and was selected because of its ease of implementation.~\cite{nesbet:311}
In the Nesbet method, a expansion coefficient $c_\mu$ is updated  by $\Delta c_\mu$
\begin{align}
	c_\mu = c_\mu + \Delta c_\mu , 
\end{align} 
where the update is calculated as
\begin{align}
	\Delta c_\mu &= 
	\frac{\sigma_\mu}{E\tilde{1}_{\mu \mu}-\tilde{H}_{\mu \mu}}, \\
	\sigma_\mu &=
	\sum_i \tilde{H}_{\mu i} c_i - E \sum_i \tilde{1}_{\mu i} c_i . 
\end{align}
The energy is updated at each step using 
\begin{align}
	\Delta E &=
	\frac{\sigma_\mu \Delta c_\mu}{D + \Delta D}, \\
	\Delta D &=
	\Delta c_\mu
	\left[
		2 \sum_i S_{\mu i} c_i 
		+ S_{\mu \mu} \Delta c_\mu 
	\right] .
\end{align}
The FCI energy can be recovered from the CT calculation by setting $G=1$
\begin{align}
	E_\mathrm{FCI}  = \lim_{G \rightarrow 1} E_\mathrm{CT}.
\end{align}
From the above relationship, we expect that the CTH energy calculated with 
$G_\mathrm{min}$ will be lower than the FCI results. In the following
section, we perform CTH calculations on well-studied two-electron systems and compare
calculated energies with reported benchmark values. 

\section{Calculations and results of benchmark systems}
\label{sec:comp_details}
\begin{figure}
\begin{center}
\includegraphics[width=3.4in]{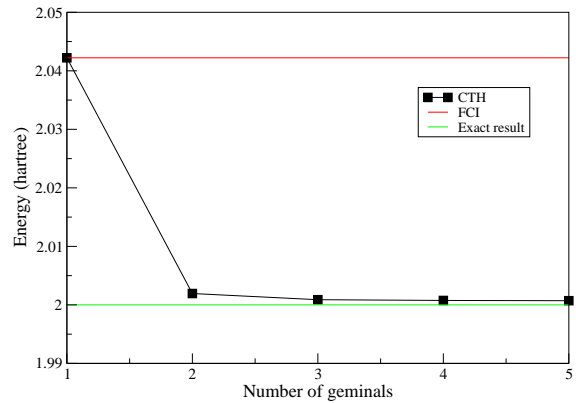}
\caption{Comparison of the exact ground state energy of Hooke's atom with the results from CTH and FCI calculations.}
\label{fig:hooke_eng}
\end{center}
\end{figure}

\begin{figure}
   \centering
  \includegraphics[width=3.4in]{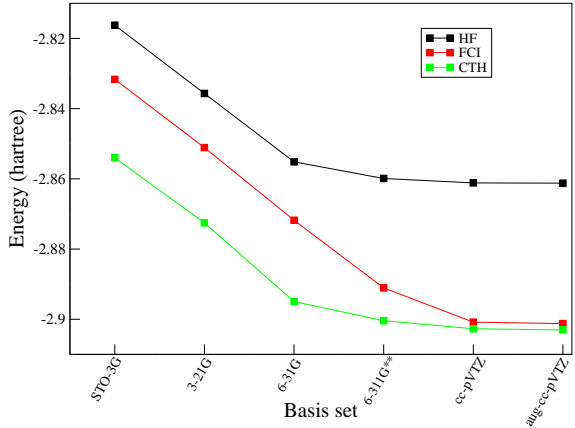}
 \vspace{-0.4cm}
  \caption{Effect of basis set on ground state energy of Helium for HF, FCI, and CTH methods.}
  \label{fig:helium_eng_comparison}
\end{figure}

\begin{figure}
\begin{center}
\includegraphics[width=3.4in]{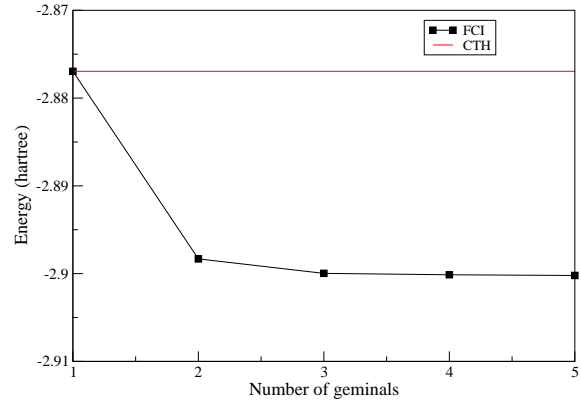}
\caption{Convergence of the CT Hamiltonian energy of the Helium atom with respect to the number of geminal functions. The calculations were performed with 6-311G basis set.}
\label{fig:helium_eng}
\end{center}
\end{figure}

The Hooke's atom is one of the few correlated two-electron systems for which the 
Schr\"{o}dinger equation can be solved analytically. This feature has made it
a testing ground for a wide variety of methods.~\cite{PhysRevA.85.032511,PhysRevA.72.062110,PhysRevA.83.042518,0253-6102-55-4-06}
The Hooke's atom consists of two electrons in a parabolic potential. 
The Hamiltonian of that system can be written as
\begin{align}
  \label{H-hooke}
  \hat{H} = -{1 \over 2} \nabla^2_1 -{1 \over 2} \nabla^2_2 + {1 \over 2}kr_1^2
  + {1 \over 2}kr_2^2 + {1 \over  r_{12}}
\end{align}
where, all the quantities are expressed in the atomic units. The interaction between an electron and the nucleus is described with the harmonic potential.
For $k=0.25\,\mathrm{a.u.}$, the Schr\"{o}dinger equation
can be solved exactly 
 and the 
ground state energy is equal to 2.0 Hartrees.~\cite{B926389F} The Hooke's
atom provides an ideal ground for testing the CTH method.
The CTH calculations were performed using the 6-311G basis and the geminal 
parameters were obtained variationally from the solution of Eq. \eqref{eq:gmin}.
 The energy was
converged with respect to number of geminal parameters $N_\mathrm{g}$, and the results 
are presented in Fig. \ref{fig:hooke_eng}. 
It is seen that the energy was converged after addition 
of four geminal parameters and the optimized geminal parameters are listed in 
Table~\ref{tab:hooke_geminal_parameter}.
\begin{table}
  \label{ref-hookes-atom}
  \begin{center}
    \caption{Geminal parameters for Hooke's atom using the 6-311G basis set}
    \label{tab:hooke_geminal_parameter}
      \begin{tabular}{ l c r } 
        \hline 
          Number &  $b_{k}$   & $\gamma_{k}$ \\
        \hline 
          1      &    1.0000  &   0.0000 \\
          2      &   -0.6090  &   0.1050 \\
          3      &   -0.0709  &   2.350  \\
          4      &    0.0216  &   0.175  \\
          5      &   -0.0132  &   1.120  \\ 
        \hline 
      \end{tabular}
    \end{center}
\end{table}
Comparing the energy with the exact result of 2.0 Hartrees, it is seen that
the $\tilde{E}[G_\mathrm{min}] $ is slightly higher by 0.770 mHartrees (or 0.483 kcal/mol).
The optimized Slater determinant $\Phi$  obtained in the previous step is used 
as the reference wavefunction for the CTH calculations and the
results are summarized in Fig. \ref{fig:hooke_eng}. For $G=1$, the CTH energy 
is identical to the FCI energy. However, inclusion of additional geminal terms
makes the CTH energy lower than the energy from the FCI calculation. It is seen that the
CTH energy is in good agreement with the exact analytical results and is higher by 
0.000296 Hartrees, these results are provided in Table ~\ref{tab:hooke_eng_comparison}. 
\begin{table}
  \begin{center}
    \caption{Difference between exact and calculated energy Hooke's atom using the CTH method}
    \label{tab:hooke_eng_comparison}
    \begin{tabular}{ c c c c c }
      \hline 
         Hartree  & kcal/mol & kJ/mol & eV      & cm$^{-1}$ \\ 
        \hline 
         0.000296 & 0.186    & 0.777  & 0.00805 & 65.0 \\
      \hline 
     \end{tabular}
   \end{center}
\end{table}
The CTH calculations were also carried out for the helium atom 
and the results are presented in Fig.~\ref{fig:helium_eng_comparison}.
The calculations were performed using
different basis functions, and the results were compared with HF and FCI values. 
It is seen that for small basis sets, the $\tilde{E}$ energy is lower than the FCI
energy. We expect this because of the inclusion of the optimized geminal terms. The key result from Fig. \ref{fig:helium_eng_comparison} is that 
for small basis sets, the CTH method provides a substantial 
lowering of energy with respect to the corresponding FCI values. The CTH 
calculations with respect to a small basis provides a wavefunction that is comparable to 
the FCI wavefunction at much larger basis functions. Since the cost of the 
FCI expansion increases sharply with the size of the underlying 1-particle basis,
the CTH method provides an appealing alternative for obtaining
accurate results when an FCI calculation is prohibitively expensive.
The dependence of the CTH energies on the number of geminal parameters
is shown in Fig.  \ref{fig:helium_eng}
 and the   
optimized geminal parameters for the helium atom are listed in Table \ref{tab:helium_geminal_param}.
\begin{table}
  \begin{center}
    \caption{Geminal parameters for Helium atom using 6-311G}
    \label{tab:helium_geminal_param}
      \begin{tabular}{c c c} 
      \hline 
        Number &  $b_k$    & $\gamma_k$ \\ 
      \hline 
        1      &  1.000000  & 0.00000 \\
        2      & -0.320260  & 0.57816 \\
        3      & -0.063365  & 10.3760 \\
        4      &  0.020918  & 0.83536 \\ 
        5      & -0.029282  & 0.08799 \\ 
      \hline 
    \end{tabular}
  \end{center}
\end{table} 

\section{Discussion and conclusions}
\label{sec:discussion_conclusions}
The first geminal parameter is always set to $b_1 = 1$ and $\gamma_1 = 0$
and is never optimized during the calculations. When all the other
$N_\mathrm{g}-1$ geminal parameters are set to zero, these values 
of $b_1$ and $\gamma_1$ represent the $G=1$ limit. Geminal parameters 
from $b_2 \dots b_{N_\mathrm{g}}$ and  $\gamma_2 \dots \gamma_{N_\mathrm{g}}$
are optimized to obtain $G_\mathrm{min}$ as described in Eq. \eqref{eq:gmin}. 
This procedure 
ensures that the optimized energy is always bounded from above by the HF energy.
Figures \ref{fig:hooke_eng} and \ref{fig:helium_eng_comparison} 
show the effect of inclusion of additional geminal 
parameters and it is seen that the second geminal parameter
lowers the energy significantly. This is an important result
and clearly indicates the importance of the geminal function 
in construction of the congruent transformed Hamiltonian.
The set of $\{b_k\}$ was optimized without any constraint and it is 
seen from Tables \ref{tab:hooke_geminal_parameter} and \ref{tab:helium_geminal_param}
that the overal geminal parameter is negative. 
This is an expected result and is in agreement with previous work 
on explicitly correlated methods.~\cite{Hofener200925,shiozaki:034113,WCMS:WCMS68} The negative values of geminal
parameters indicate the role of the geminal function in 
providing a better description of the Coulomb hole. 

The analytical forms of the GTG functions are inherently approximate and are 
not capable of describing the cusp correctly because their first derivative vanishes
in the limit of $r_\mathrm{ee}=0$
\begin{align}
	\left( \frac{\partial G}{\partial r_{\mathrm{ee}}} \right)_{r_\mathrm{ee}=0} = 0.
\end{align} 
To assess the quality of the CTH energy, it is important 
to estimate how much of an error this feature introduces in the
calculated energy. For the Hooke's atom this can be done in a 
a straightforward manner since the analytical solution of the Schr\"{o}dinger
equation is known.  From Table  \ref{tab:hooke_eng_comparison}, it is seen that the CTH 
energy is close to the exact ground state energy and is higher by 0.296 mHartrees 
or  0.186 kcal/mol. This difference between the CTH and the exact energy 
represents the upper bound in the error that one can expect for this system by 
approximating the cusp with GTG functions. For the helium atom, the 
situation is less straightforward because we do not have access to the exact 
solution. Instead, we compared the CTH energies
with other high-level methods from previous
studies ~\cite{hylleraas,RefWorks:216,RefWorks:217,RefWorks:201,RefWorks:202,RefWorks:213,RefWorks:208,RefWorks:212,RefWorks:214} that include
the exact cusp condition in the wavefunction. 
In order to achieve the best CTH energy, the
calculation was performed with an aug-cc-pVTZ basis set and geminal parameters were optimized with respect to the aug-cc-pVTZ basis.
Comparing the CTH method with the highly accurate ICI method by Nakatsuji~\cite{RefWorks:201,RefWorks:202}, 
it is seen that CTH energy is higher than the ICI energy by 0.429 kcal/mol. The comparison of the CTH calculation to the ICI method
and other highly accurate results can be seen in Table ~\ref{tab:helium_eng_comparison}. 
\begin{table} 
  \caption{Comparison of ground state energy (in Hartrees) of the helium atom}
   \label{tab:helium_eng_comparison}
    \begin{center}
      \begin{tabular}{l c c c c}
      \hline 
        $E$ & function & ref \\ 
      \hline
               -2.900233        & FCI  & \cite{comp_chem_data} \\ 
               -2.903041       & CTH & this work \\ 
               -2.903724 & free ICI & \cite{RefWorks:201,RefWorks:202} \\
               % -2.9037243770341195983115922451944044466969 & free ICI &
     \hline 
   \end{tabular}
  \end{center}
\end{table}  
The impact of electron-electron cusp on ground state energy was investigated 
in detail by Prendergast~\cite{prendergast:1626} and coworker using CI and QMC methods. 
Their study concluded that one can still expect to get mHartree level of 
accuracy even in situations where the exact cusp condition is not satisfied. 
Our study using GTG functions also confirms this observation.
The use of GTG functions in the CTH method, represents a trade-off 
between the implementation of exact cusp condition and analytical expression for 
 computing the Gaussian-type geminal integrals. 

In conclusion, the congruent transformation of the electronic Hamiltonian
using Gaussian type geminal function is presented as a general method for
calculating accurate ground state energy. 
 The form of the congruent transformed Hamiltonian can be systematically improved
by using the geminal function. It was found that a small number of 
geminal functions are needed to converge the energy. Futhermore, addition of just one geminal parameter results in a substantial improvement in the 
accuracy of the wave function. 
For a given finite basis set
the CTH energy was found to be lower that the FCI calculation on untransformed Hamiltonian.
The results indicate that the congruent transformed
Hamiltonian provides a viable alternative for obtaining FCI quality energy using a 
smaller underlying 1-particle basis set.

\begin{acknowledgments}
We gratefully acknowledge the support from 
Syracuse University, Graz University of Technology, NSF REU program
Grant No. CHE-0850756, and NSF iREU program
Grant No. CHE-0755383 for this work.
\end{acknowledgments}

%\section*{Acknowledgement}

%%%%%%%%%%%%%%%%%%%%%%%%%%%%%%%%%%%%%%%%%%%%%%%%%%%%%%%%%%%%%%%%%%%%%%%%%%%%%%%%
%%\bibliographystyle{groupjcp}
\bibliography{ref_version_1_he}

\newpage
%\clearpage

\clearpage

\newpage    
%\clearpage

\newpage

\clearpage

\clearpage

\newpage
%\clearpage

\clearpage

\end{document}